\documentstyle[aps,preprint]{revtex}
\begin{document}
\title{Gravitational Waves from The Newtonian plus H\'enon-Heiles 
System\thanks{Accepted for publication in {\it Phys.Lett.A}}}
\author{Fernando Kokubun}
\address{Departamento de F\'{\i}sica, Funda\c{c}\~ao Universidade do Rio\\
Grande \\
Caixa Postal 474, CEP 96201-900, Rio Grande, RS, Brazil}
\maketitle

\begin{abstract}
In this work we analyze the emission of gravitational waves from a
gravitational system described by a Newtonian term plus a  H\'enon-Heiles
term. The main concern is to analyze how the inclusion of the Newtonian term
changes the emission of gravitational waves, considering its emission in the
chaotic and regular regime.
\end{abstract}


In a previous paper we analyzed the emission of gravitational waves from a
H\'{e}non-Heiles System, showing the qualitative differences between
gravitational waves emission from chaotic and regular system \cite{prd98}.
Although interesting, the H\'{e}non-Heiles system is not realistic and it
was originally intended to study galactic dynamics, such that the emission
of gravitational waves is expected to be negligible. Thus in this work we
include a Newtonian term $\propto r^{-1}$ to the H\'{e}non-Heiles potential.
This class of potential was analysed by Vieira and Letelier \cite{werner},
and correspond to a black-hole with external halo. This type of system is
more realistic than simple H\'{e}non-Heiles system and its gravitational
waves emission may be important due to its stellar system size.

In this work, the dynamics are determined by the gravitational potential 
$V=-GMm/r+m\omega ^2[(x^2+y^2)/2+(x^2y-y^3/3)/a].$ Using the usual approach,
we work in a dimensionless system, such that the gravitational potential is
reduced to 
\begin{equation}
\Phi =-\frac \alpha {\tilde{r}}+\left[ \frac{\eta ^2+\xi ^2}2+\eta ^2\xi -
\frac{\xi ^3}3\right]   \label{eq:pot}
\end{equation}
with $\Phi \equiv V/(m\omega ^2a^2),\;\alpha \equiv GM/(\omega ^2a^3),\tilde{
r}\equiv r/a\;,\eta \equiv x/a$ and $\xi =y/a$. A main difference between
pure H\'{e}non-Heiles \cite{henon} and above potential, is that bounded
trajectories in the first case are limited to the range $0<E<1/6$, whereas
in our case we have bounded trajectories even with $E<0$. Other difference
is that the upper limit in the energy ( $E_{max}=1/6\sim 0.167$) may be
lower than a pure H\'{e}non-Heiles case, its value depending of the $\alpha $
parameter. For example with $\alpha =0.1$, we have $E_{max}\sim 0.07$ and
with $\alpha =0.01$ we have $E_{max}\sim 0.156$. With above potential, we
calculated the trajectories using numerical simulations, changing energy $E$, 
the parameter $\alpha $ and initial conditions. For our simulations 
we used $\alpha=0.01,0.001\;{\rm and}\;0.0001$. The main reason to these
 choice is to
maintain some resemblance with H\'{e}non-Heiles case. (Later in the text, we
discuss the physical meaning and limitations of this choice.) Note that with
these $\alpha $ we have always $E_{max}\sim 1/6$. We used in our simulations
three distinct values for the energies, i.e., $E=0.1,0.01$ and $0.001$. In
this range of energy we obtain a good mixture of chaotic and regular motions
for our choice of $\alpha $s, permitting a better comparison between
emission of gravitational waves from both cases.

The power emitted as a gravitational waves was determined using the standard
quadrupole formula 
\begin{equation}
P\equiv\frac{dE}{dt}=\lambda \stackrel{...}{q}_{\alpha\beta}^2
\end{equation}
where 
\begin{eqnarray}
\lambda&=&\frac{Gm}{45 c^5}a^2\omega^3
\end{eqnarray}
and quadrupole terms are given by $q_{\eta\eta}=2\eta^2-\xi^2\;,q_{\xi\xi}=2
\xi^2-\eta^2$ and $q_{\eta\xi}=3\eta\xi$. (In what follow we use $T$ instead
of $\omega$, with $T\equiv 2\pi/\omega.$)

In the figures \ref{fig:1},\ref{fig:1b} and \ref{fig:2} we show some results
of our simulations. We plot trajectories in the X-Y plane, Poincar\'{e}
section and the spectrum of the emitted gravitational waves for $\alpha =0.01
$ and three different energies $E=0.10,0.01$ and $0.001$. In each case we
plot one chaotic and one regular motion. The maximum power emitted as a
gravitational waves in each case are given in the table \ref{tab:0}. Note
that in all cases, except when $\alpha =0.01$ and $E=0.1$, the power emitted
in chaotic case is greater than regular case (
This is an exception, as can be seen  in the table \ref{tab:ap3} where we
show a more detailed statistics of  emitted power.).

The irregular characteristic of the gravitational waves spectrum is show in
all chaotic case, being more strong with $E=0.1$. In particular note the
strong differences of the emitted spectrum of the chaotic and regular case
in the figure \ref{fig:1}.

In the tables \ref{tab:ap1},\ref{tab:ap2} and \ref{tab:ap3} we show with
more details the power emitted as a gravitational waves for several $\alpha$
and $E$. As expected, the emission in the chaotic case is greater than
regular case \cite{prd98}. However, for lower $E$ the differences between
regular and chaotic case becomes negligible. This happens because for high
 $E$, the trajectories reach the boundaries of  the H\'enon-Heiles term. In
this case the trajectories of the particles  suffer a strong changes, 
resulting in a higher emission of gravitational waves. On the other hand for
lower $E$, the trajectories are confined far away from H\'enon-Heiles
boundaries, such that it not suffer strong changes as the  previous case.
However, in all cases the spectrum of the emitted gravitational waves are
much more irregular in the chaotic case when compared with the regular case.

As soon the potential which we are using is more realistic than a simple
H\'{e}non-Heiles case, it will be interesting to verify in which conditions
the emission of gravitational waves will be important. To do it, it is
important to fix the values of $\alpha $ and $\lambda $. Thus in order to
obtain a better understanding of $\alpha $ and $\lambda $, we assume two
typical size for the system: (1) Solar System and (2) binary pulsar
(typically PSR 1913+16) with parameters $(a_1,T_1),(a_2,T_2)$ respectively.
Then we set $T=T_oT_i\;,a=a_oa_i\;,i=1,2$ with $a_1=1.5\times 10^{11}m$, $
T_1=3.1\times 10^7s$, $a_2=2.0\times 10^9m$ and $T_2=2.8\times 10^4m$ and $
M=\beta _1M_{\odot },m=\beta _2M_{\odot }$ with $M_{\odot }=2.0\times
10^{30}g$ (Solar mass). Thus 
\begin{eqnarray}
\mbox{Solar System size}:\; &&\alpha _1=0.96\beta _1\frac{T_o^2}{a_o^3}
\;;\;\lambda _1=2.29\times 10^{-22}\beta _2\frac{a_o^2}{T_o^3} \\
\mbox{Binary Sistem size}:\; &&\alpha _2=0.33\beta _1\frac{T_o^2}{a_o^3}
\;;\;\lambda _2=5.5\times 10^{-17}\beta _2\frac{a_o^2}{T_o^3}
\end{eqnarray}

In the tables \ref{tab:at1} and \ref{tab:at2} we show some numerical values
for $a_o,T_o$ and $\lambda $ using $\alpha =0.0001,0.001$ and $0.01$ and $
\beta _1=\beta _2=1.4$. Also, we show $\Delta E/E_i\sim (<P_{max}>T_o)/E_i$
which is the energy lost as a gravitational waves during a typical crossing
time $T_o$ and we use two values for $E_i$, i.e., $E_i=E_{max}=0.10$ and $
E_i=E_{min}=0.001$. This last column show how much the emission are
important to the dynamics of the system. Only when $\Delta E/E_o\ll 1$ the
emission of gravitational waves will be negligible to the dynamics of the
system. (Although $\Delta E/E_o>1$ is meaningless, we show these values only
for the sake of completeness.) From the tables \ref{tab:at1} and \ref
{tab:at2} we see that when the system size is of the order of a binary
pulsar the emission of gravitational waves will be important. On the other
hand when the size is of the order of the Solar System the energy lost as a
gravitational waves are negligible only in the cases (Ia) and (Ib). In all
other cases the energy lost as a gravitational waves are of the same order
of magnitude than its initial energy.

Note that $\Delta E/E$ is lower when $\alpha$ and $E_o$ are lower. 
As said  before, in these
cases the size of typical trajectory is small compared with the size of
external halo. For example when $\alpha=0.0001$ and energies $E_o=0.001$ and 
$E=0.10 $ we obtained with numerical simulations that $|\xi|_{max}\sim 0.07$
, and $|\xi|_{max}\sim 0.6$ respectively. These result show that  the size
of the trajectories with  small energy are smaller than trajectories  with
high energy (as expected). This mean that in the first case (small energy)
the trajectories do not reach the boundaries defined by the H\'enon-Heiles
term, whereas in the high energy case these boundaries  can be reached. Thus
in the first case (low energy), the external halo is felt only as a small
perturbations (note that for small $(\eta,\xi)$ the H\'enon-Heiles term is
negligible in  respect to the Newtonian term).

Therefore a potential like as given by equation \ref{eq:pot}, is useful only
when external halo is felt as a small perturbations (as soon we consider the
effect of the gravitational waves to the dynamics of the system).

However, even with inclusion of the Newtonian term, the emission of
gravitational waves are greater in the chaotic case than regular case --
except for low energy case -- , like our previous results \cite{prd98}. Also
the spectrum of the emitted wave is highly irregular, thus we can speculate
that with a careful analyze of the gravitational waves spectrum we can
detect a gravitational chaotic system. But it is important to note that our
results show that if a system is described by a Newtonian potential with a
H\'{e}non-Heiles term its typical size need to be greater than a Solar
System size. If this is not the case, the emission of gravitational waves
will be important and it will collapse quickly. This show that although more
realistic, a single inclusion of the Newtonian term is not sufficient as a
realistic core-halo system model (see for example \cite{vieira2} for a more
realistic approach).

This work was partially supported by Funda\c {c}\~{a}o de Amparo a Pesquisa
do Rio Grande do Sul (FAPERGS), brazilian financial agency.

\begin{figure}[tbp]
\caption{An examples of chaotic and regular motions with $E=10^{-1}$ and $
\alpha=10^{-2}$. The left box show the trajectories in the X-Y plane, the
central box the Poincar\'e section and the right box the some portion of the
gravitaional waves spectrum. The chaotic case is show in the top figures
set, and regular case at bottom set.}
\label{fig:1}
\end{figure}

\begin{figure}[tbp]
\caption{Same as previous figure, but now using $E=10^{-2}$.}
\label{fig:1b}
\end{figure}

\begin{figure}[tbp]
\caption{Same as previous figure, but now using $E=10^{-3}$. }
\label{fig:2}
\end{figure}

\begin{table}[tbp]
\begin{tabular}{ccc}
$E$ & $P_{max}/\lambda$ &  \\ \hline
0.100 & $5.21\times 10^{17}$ & chaotic \\ 
0.100 & $5.34\times 10^{17}$ & regular \\ 
0.010 & $1.49\times 10^{16}$ & chaotic \\ 
0.010 & $6.73\times 10^{15}$ & regular \\ 
0.001 & $1.04\times 10^{16}$ & chaotic \\ 
0.001 & $5.23\times 10^{15}$ & regular \\ \hline
\end{tabular}
\caption{Maximum emiited power as a gravitational waves for the cases shown
in the figures \ref{fig:1},\ref{fig:1b} and \ref{fig:2}}
\label{tab:0}
\end{table}

\begin{table}[tbp]
\begin{tabular}{cccc}
$\alpha$ & E & $P_{max}/\lambda $ & Note \\ \hline
0.0001 & 0.100 & $(8.15\pm 3.78)\times 10^{12}$ & chaotic+regular \\ 
0.0001 & 0.100 & $(5.89\pm 2.49)\times 10^{12}$ & regular \\ 
0.0001 & 0.100 & $(1.00\pm 0.36)\times 10^{13}$ & chaotic \\ 
0.0001 & 0.010 & $(6.26\pm 2.74)\times 10^{10}$ & chaotic+regular \\ 
0.0001 & 0.010 & $(6.43\pm 2.81)\times 10^{10}$ & regular \\ 
0.0001 & 0.010 & $(5.65\pm 2.43)\times 10^{10}$ & chaotic \\ 
0.0001 & 0.001 & $(2.72\pm 1.22)\times 10^{9} $ & regular \\ \hline
\end{tabular}
\caption{Maximum power emitted as a gravitationala waves with $\alpha=0.0001$
for three energies $E=0.1,0.01$ and $0.001$ }
\label{tab:ap1}
\end{table}

\begin{table}[tbp]
\begin{tabular}{cccc}
$\alpha$ & E & $P_{max}/\lambda $ & Note \\ \hline
0.0010 & 0.100 & $(1.53\pm 0.73)\times 10^{17}$ & chaotic+regular \\ 
0.0010 & 0.100 & $(9.61\pm 5.44)\times 10^{16}$ & regular \\ 
0.0010 & 0.100 & $(1.78\pm 0.66)\times 10^{17}$ & chaotic \\ 
0.0010 & 0.010 & $(2.09\pm 0.96)\times 10^{15}$ & chaotic+regular \\ 
0.0010 & 0.010 & $(1.25\pm 0.45)\times 10^{15}$ & regular \\ 
0.0010 & 0.010 & $(2.31\pm 0.94)\times 10^{15}$ & chaotic \\ 
0.0010 & 0.001 & $(4.97\pm 2.41)\times 10^{14}$ & chaotic+regular \\ 
0.0010 & 0.001 & $(5.00\pm 2.60)\times 10^{14}$ & regular \\ 
0.0010 & 0.001 & $(4.87\pm 1.70)\times 10^{14}$ & chaotic \\ \hline
\end{tabular}
\caption{Maximum power emitted as a gravitationala waves with $\alpha=0.001$
for three energies $E=0.1,0.01$ and $0.001$ }
\label{tab:ap2}
\end{table}

\begin{table}[tbp]
\begin{tabular}{cccc}
$\alpha$ & E & $P_{max}/\lambda $ & Note \\ \hline
0.0100 & 0.100 & $(2.55\pm 0.79)\times 10^{17}$ & chaotic+regular \\ 
0.0100 & 0.100 & $(1.41\pm 0.59)\times 10^{17}$ & regular \\ 
0.0100 & 0.100 & $(2.60\pm 0.76)\times 10^{17}$ & chaotic \\ 
0.0100 & 0.010 & $(1.76\pm 0.82)\times 10^{16}$ & chaotic+regular \\ 
0.0100 & 0.010 & $(1.38\pm 0.62)\times 10^{16}$ & regular \\ 
0.0100 & 0.010 & $(1.83\pm 0.84)\times 10^{16}$ & chaotic \\ 
0.0100 & 0.001 & $(1.07\pm 0.57)\times 10^{16}$ & chaotic+regular \\ 
0.0100 & 0.001 & $(8.69\pm 3.80)\times 10^{15}$ & regular \\ 
0.0100 & 0.001 & $(1.56\pm 0.65)\times 10^{16}$ & chaotic \\ \hline
\end{tabular}
\caption{Maximum power emitted as a gravitationala waves with $\alpha=0.01$
for three energies $E=0.1,0.01$ and $0.001$.}
\label{tab:ap3}
\end{table}

\begin{table}[tbp]
\begin{tabular}{ccccccc}
& $\alpha$ & $a_o$ & $T_o$ & $\lambda$ & $\Delta E/E_{max}$ & $\Delta
E/E_{min}$ \\ \hline
(Ia) & 0.0001 & $1$ & $8.63\times 10^{-3}$ & $4.99\times 10^{-16}$ & $
4.32\times 10^{-4}$ & $1.21\times 10^{-5}$ \\ 
(Ib) & 0.0001 & $23.7$ & $1$ & $1.30\times 10^{-19}$ & $5.0\times 10^{-5}$ & 
$3.5 \times 10^{-7}$ \\ 
(Ic) & 0.0010 & $1$ & $3.23\times 10^{-2}$ & $1.21\times 10^{-17}$ & $
6.94\times 10^{-1}$ & $1.87\times 10^{-1}$ \\ 
(Id) & 0.0010 & $9.80$ & $1$ & $2.80\times 10^{-20}$ & $4.98\times 10^{-2}$
& $1.34\times 10^{-2}$ \\ 
(Ie) & 0.0100 & $1$ & $8.63\times 10^{-2}$ & $4.99\times 10^{-19}$ & $
1.12\times 10^{-1}$ & $4.31\times 10^{-1}$ \\ 
(If) & 0.0100 & $5.12$ & $1$ & $8.39\times 10^{-21}$ & $2.18\times 10^{-2}$
& $8.39\times 10^{-2}$ \\ \hline
\end{tabular}
\caption{Some set of values for $a_o,T_o$ and $\lambda$ when typical size
correspond to our Solar System with $a_1=1.5\times 10^{11}\;m$ and $
T_2=3.1\times 10^{7}\;s$ (see text). In the last two columns, we uses $
E_{max}=0.10$ and $E_{min}=0.001$ }
\label{tab:at1}
\end{table}

\begin{table}[tbp]
\begin{tabular}{ccccccc}
& $\alpha$ & $a_o$ & $T_o$ & $\lambda$ & $\Delta E/E_{max}$ & $\Delta
E/E_{min}$ \\ \hline
(IIa) & 0.0001 & $1$ & $1.47\times 10^{-2}$ & $2.42\times 10^{-11}$ & $
3.56\times10^{1}$ & $9.67\times 10^{-1}$ \\ 
(IIb) & 0.0001 & $16.7$ & $1$ & $2.12\times10^{-14}$ & $2.12$ & $5.77\times
10^{-2}$ \\ 
(IIc) & 0.0010 & $1$ & $5.51\times 10^{-2}$ & $3.31\times 10^{-13}$ & $
3.25\times 10^3$ & $8.76\times 10^{3}$ \\ 
(IId) & 0.0010 & 6.91 & 1 & $2.63\times 10^{-15}$ & $4.68\times 10^{3}$ & $
1.26\times 10^{3}$ \\ 
(IIe) & 0.0100 & $1$ & $0.14$ & $2.45\times 10^{-14}$ & $4.75\times 10^5$ & $
1.7\times 10^6$ \\ 
(IIf) & 0.0100 & $3.59$ & $1$ & $9.92\times 10^{-16}$ & $2.6\times 10^3$ & $
1.0\times 10^4$ \\ \hline
\end{tabular}
\caption{Some set of values for $a_o,T_o$ and $\lambda$ when typical size
correspond to a bynary pulsar with $a_2=2.0\times 10^9\;m$ and $
T_2=2.8\times 10^4\;s$ (see text). In the last two columns, we uses $
E_{max}=0.10$ and $E_{min}=0.001$ }
\label{tab:at2}
\end{table}
\end{document}